\documentclass[aps,twocolumn,prl,showpacs,superscriptaddress]{revtex4}
\usepackage{graphicx}
\usepackage{dcolumn}
\usepackage{bm}
\usepackage{epsfig}
\usepackage{pstcol,pst-grad}

\begin{document}

\preprint{ECM}

\title{Acoustic emission across the magnetostructural transition of the
  giant magnetocaloric Gd$_5$Si$_2$Ge$_2$ compound}

\author{Francisco-Jos\'e P\'erez-Reche}

\affiliation{ Departament d'Estructura i Constituents de la Mat\`eria,
  Universitat de Barcelona \\ Diagonal 647, Facultat de F\'{\i}sica,
  08028 Barcelona, Catalonia}
\author{F\`elix Casanova}
\affiliation{Departament de F\'{\i}sica Fonamental. Universitat de
  Barcelona \\ Diagonal 647, Facultat de F\'{\i}sica, 08028 Barcelona,
  Catalonia}
\author{Eduard Vives}
\author{Llu\'{\i}s Ma\~{n}osa}
\email{lluis@ecm.ub.es}
\author{Antoni Planes}
\author{Jordi Marcos}

\affiliation{ Departament d'Estructura i Constituents de la Mat\`eria,
  Universitat de Barcelona \\ Diagonal 647, Facultat de F\'{\i}sica,
  08028 Barcelona, Catalonia}

\author{Xavier Batlle}
\author{Am\'{\i}lcar Labarta}
\affiliation{Departament de F\'{\i}sica Fonamental. Universitat de
  Barcelona \\ Diagonal 647, Facultat de F\'{\i}sica, 08028 Barcelona,
  Catalonia}

\date{\today}

\begin{abstract}
  We report on the existence of acoustic emission during the
  paramagnetic-monoclinic $\leftrightarrow$ ferromagnetic-orthorhombic
  magnetostructural phase transition in the giant magnetocaloric
  Gd$_5$Si$_2$Ge$_2$ compound. The transition kinetics have been
  analyzed from the detected acoustic signals. It is shown that
  this transition proceeds by avalanches between metastable
  states.
\end{abstract}


\maketitle

\section{INTRODUCTION}

The interplay of magnetism and structure is at the origin of many
of the technologically important properties of functional
materials such as the giant magnetocaloric effect, magnetic shape
memory and giant magnetoresistance \cite{Planes2005}.
Magnetocaloric effects associated with magnetic transitions have
received a considerable amount of interest in the recent years
\cite{Tegus2002,Gschneidner2005}.   Gd$_5$(Si$_x$Ge$_{1-x}$)$_4$
intermetallics are prominent among the materials exhibiting such
effect and have received a great deal of attention
\cite{Pecharsky1997,Provenzano2004}. In these compounds, the giant
magnetocaloric effect is due to the occurrence of a magnetic phase
transition which also involves a crystallographic structural
change \cite{Choe2000,Morellon1998}. This transition is first
order, reversible and can be induced either by changing the
temperature \cite{Morellon1998,Choe2000}, the pressure
\cite{Morellon2004,Mudryk2005}, or by application of a magnetic
field \cite{Morellon1998,Casanova2004}. In the range $0.24 \leq x
\leq 0.5$, the transition goes from a paramagnetic monoclinic
phase towards a ferromagnetic orthorhombic phase on cooling
\cite{Morellon2000}.

A structural change in a material usually produces changes in the
internal strain field, which give rise to elastic waves in the
ultrasonic range propagating within the material.  These elastic
waves are known as Acoustic Emission (AE) and they convey
information on the dynamics of the mechanism that has generated
them \cite{Manosa1989}. Prototypical solid-solid phase transitions
with associated AE are martensitic transitions, for which the
analysis of the AE has provided valuable information on the
mechanisms and kinetics of the transition
\cite{Yu1987,Manosa1989,Manosa1990,Vives1994a,PerezReche2001}.

The mechanisms of the crystallographic change in
Gd$_5$(Si$_x$Ge$_{1-x}$)$_4$ involve shearing of planes perpendicular
to the long $b$ axis \cite{Choe2000,Meyers2003}.  Since this mechanism
shares some similarities with martensitic transitions, it has been
termed ``martensitic-like''.  Hence, it is expected that such a
structural change can generate AE in these kinds of materials.  In the
present paper we show the existence of AE during the magnetostructural
transition of Gd$_5$Si$_2$Ge$_2$ and we analyze the kinetics of the
transition from the detected AE.  Results are discussed in comparison
with the well-established AE results in thermoelastic martensitic
transitions.

\section{EXPERIMENTAL DETAILS}

A Gd$_5$Si$_2$Ge$_2$ sample was synthesized by arc melting the
pure elements (commercial 99.9 wt \% Gd, 99.9999 wt \% Si and
99.999 wt \% Ge) in the appropriate stoichiometry under a
high-purity argon atmosphere. The sample was placed in a
water-cooled copper crucible and was melted several times to
ensure good homogeneity. From the as-prepared buttons, the sample
was cut with two parallel faces and it was heat treated for
homogenization for up to 9~h at 950~$^{\circ}$C under
$10^{-5}$~mbar, inside a quartz tube in an electrical resistance
furnace.  After annealing, the quartz tube was quickly taken out
of the furnace and was left to cool to room temperature. The
crystallographic structure of the sample was studied by
room-temperature x-ray diffraction. The material displayed the
expected monoclinic structure ($P112_1/a$) with unit cell
parametes a=7.577(1), b=14.790(3), c=7.779(1), $\gamma$=93.09(1),
in agreement with \cite{Pecharsky1997b}. Some amounts of a
secondary orthorhombic phase ($Pnma$) were also present in the
sample.

AE signals were detected by a resonant piezoelectric transducer
acoustically coupled to the surface of the sample.  The transition
was thermally induced using the experimental setup described in
Ref.~\onlinecite{PerezReche2001}.  The relative oscillations of
the sample temperature were less than 0.01~\%.  The amplified
signal (gain 62~dB) was simultaneously processed by two different
methods. On the one hand, bursts with amplitudes exceeding a fixed
threshold were stored using a digitizing oscilloscope, which is
capable of recording 10$^4$ AE pulses (1000 points per signal at
1~MHz) during the magnetostructural transition.  On the other
hand, the signal was input into a ring-down frequency meter which
renders the count rate $\dot{N}= dN/dt$ (number of signals
recorded during 1~s).

\section{RESULTS AND DISCUSSION}

\begin{figure}
\begin{center}
  \epsfig{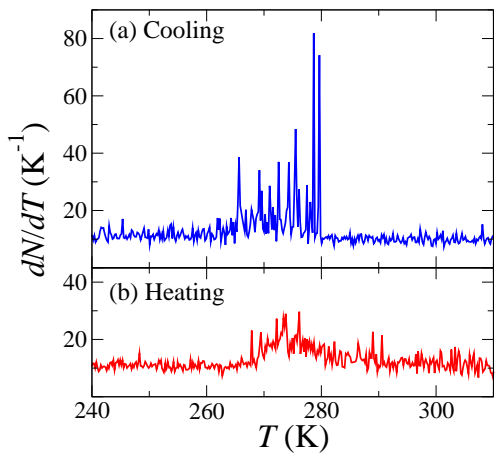}
\caption{\label{Fig1}  Reduced acoustic  activity recorded  during the
  13th cycle on (a) cooling and (b) heating.}
\end{center}
\end{figure}

Fig.~\ref{Fig1} shows an example of the reduced counting rate
($r=dN/dT=\dot{N}/\dot{T})$ recorded as a function of temperature.
A clear increase in AE activity is observed for the 280 -- 260~K
temperature range on cooling and 265 -- 290~K on heating.  These
ranges coincide with those where the magnetostructural transition
takes place, as determined from calorimetric measurements done on
the same sample, and thus confirm that there is AE generated
during the transition. Acoustic activity across the monoclinic to
orthorhombic (cooling) transition is higher than across the
reverse transition. Differences in the kinetics of the transition
between cooling and heating were also reported in
Gd$_5$Si$_{1.9}$Ge$_{2.05}$ from voltage generation measurements
\cite{Levin2001b}. It is worth noting that the recorded AE is very
weak during both, heating and cooling runs, as compared, for
instance, to the activity recorded during thermoelastic
martensitic transitions for which the overall number of recorded
counts typically exceeds the AE recorded here by more than three
orders of magnitude\cite{Manosa1989b}.

The reproducibility of the AE pattern during the magnetostructural
transition was studied.  To this end, we performed up to $\sim 30$
complete thermal cycles.  A quantitative evaluation of the similarity
in the AE pattern is achieved by computing the correlation function
$\rho_{n,n+1}$ between the reduced AE activity of two consecutive
cycles defined as \cite{PerezReche2004}:
\begin{widetext}
\begin{equation}
\rho_{n,n+1}=%
\frac{\int_{0}^{1} r_nr_{n+1}d\tau -\int_{0}^{1}
  r_n d\tau \int_{0}^{1} r_{n+1}d\tau'}%
{\sqrt{\left[\int_{0}^{1} r_n^2 d\tau -\left(
       \int_{0}^{1}
        r_n d\tau \right)^2\right]\left[\int_{0}^{1} r_{n+1}^2 d\tau
- \left(\int_{0}^{1}
        r_{n+1} d\tau \right)^2\right]}},
\end{equation}
\end{widetext}
where $\tau \equiv (T-T_i)/(T_f-T_i)$, $T_i$ is a temperature above (below) the
starting transition temperature on cooling (heating), and $T_f$ is a
temperature below (above) the finishing transition temperature on cooling
(heating). The value of $\rho_{n,n+1}$ quantifies
how much the reduced acoustic activity as a function of temperature in
the $n$th cycle resembles that of the $(n+1)$th.  When $r_n$ is very
similar to $r_{n+1}$, $\rho_{n,n+1}$ is close to unity.

\begin{figure}
\begin{center}
  \epsfig{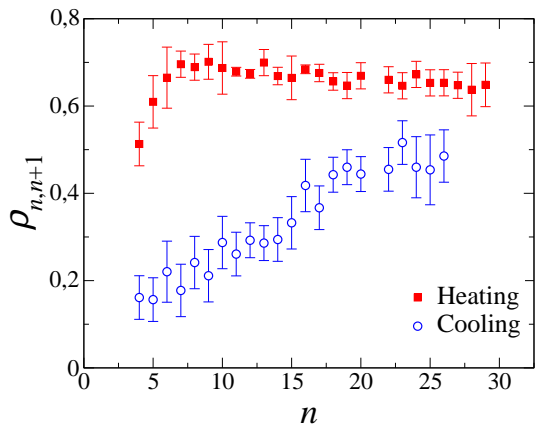}
\caption{\label{Fig2}  Correlation function  between  reduced acoustic
  activity of consecutive cycles.  Data corresponding to cooling and
  heating runs are represented by different symbols as indicated by
  the legend.}
\end{center}
\end{figure}

The correlation function $\rho_{n,n+1}$ versus the cycle number $n$ is
presented in Fig.~\ref{Fig2} for cooling and heating runs.  In both
cases, a clear increase in correlation is observed with cycling, which
indicates that the transition becomes progressively more reproducible.
Such an increase is very fast for the reverse transition (heating).
It occurs for the first 7-8 cycles, after which it saturates.  In
contrast, the increase is smoother on cooling.

\begin{figure}
\begin{center}
  \epsfig{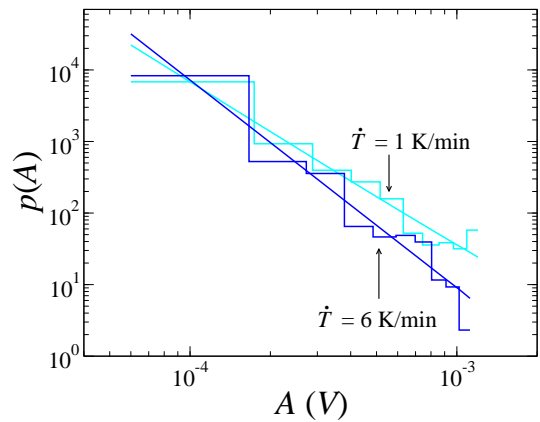}
\caption{\label{Fig3} Normalized histograms  corresponding to the distribution
  of amplitudes $p(A)$ of AE signals recorded during cooling runs at
  two different cooling rates $\dot{T}=1$~K/min and $\dot{T}=6$~K/min.
  Straight lines indicate the fits to the histograms.}
\end{center}
\end{figure}

Acoustic activity during the magnetostructural transition
corresponds to individual AE signals (bursts) which is a signature
of the jerky character of the transition.  We also performed a
statistical analysis of the individual signals recorded during
cooling. These runs are carried out once the system has been
cycled a sufficient number of times for the transition to be
reproducible.  The low number of signals recorded during heating
prevented us from performing a reliable statistical analysis of
the reverse transition. Fig.~\ref{Fig3} shows the AE amplitude
distribution $p(A)$ of the signals recorded at cooling rates
($\dot{T}$) 1~K/min and 6~K/min.  In addition, to improve
statistics, we used all the signals recorded during 40 and 170
cycles for $\dot{T}=1$~K/min and $\dot{T}=6$~K/min, respectively.
The data exhibit an apparent power-law behaviour which evidences
the absence of a characteristic scale. A more quantitative
analysis can be performed by fitting the following probability
density with two free parameters, the exponent $\alpha$ and the
exponential correction $\lambda$.
\begin{equation}
p(A)=\frac{e^{-\lambda A}
A^{-\alpha}}{\int_{A_{min}}^{A_{max}}e^{-\lambda A} A^{-\alpha}dA}
\end{equation}
where $A$ is the amplitude of the signals. The normalization
factor in the denominator corresponds to the integral of the
proposed distribution from the minimum to the maximum of the
acquired amplitudes: $A_{min}=6 \;10^{-5} V$ and $A_{max}=1.123 \;
10^{-3} V$. The estimation of the parameters $\alpha$ and
$\lambda$ is done by using the maximum likelihood method
\cite{Goldstein2004,PerezReche2004}.  This method is much more
reliable than the standard least squares method since it is not
based on the computation of the histograms which usually depend on
the binning choice and in addition it forces the fitted
probability density to be normalized.

The obtained values are $\alpha = 2.33 \pm 0.03$ and $\lambda=-130
\pm 120$ V$^{-1}$ for $\dot{T} = 1$ K/min and $\alpha=2.90 \pm
0.05$ and $\lambda=0 \pm 246$ V$^{-1}$ for $\dot{T} = 6$ K/min.
The value of $\lambda$ is compatible with $\lambda=0$ in both
cases, which indicates that the distribution is well described by
a power-law.

The magnetostructural transition in Gd$_5$Si$_2$Ge$_2$ has been
compared by several authors with martensitic transitions which
also occur via a shear mechanism.  As a particular example, the
shear distortion observed in thermoelastic martensitic transitions
in Cu-based alloys is about 0.12 \cite{Guenin1982}.  Such a shear
distortion is larger than that involved in the magnetostructural
transition in Gd$_5$Si$_2$Ge$_2$, which is $\sim 0.03$
\cite{Meyers2003}.  Such a difference may be considered as a
possible reason for the lower acoustic activity detected across
the magnetostructural transition.

There is a fundamental difference between a martensitic transition
and the magnetostructural transition in Gd$_5$Si$_2$Ge$_2$.  From
a purely crystallographic point of view the martensitic transition
occurs, on cooling, from a high symmetry phase to a low symmetry
phase.  In contrast, the transition in Gd$_5$Si$_2$Ge$_2$ occurs
from a low symmetry phase (monoclinic) to a high symmetry phase
(orthorhombic). Such a symmetry change does not fit the common
framework of phase transitions since it is expected that the high
symmetry phase corresponds to the high temperature phase.
Nevertheless, the transition in Gd$_5$Si$_2$Ge$_2$ also involves a
symmetry change associated with magnetic degrees of freedom from a
paramagnetic phase (high symmetry) to a ferromagnetic phase (low
symmetry). Hence, symmetry arguments indicate that the main
driving force for the magnetostructural phase transition arises
from the magnetic degrees of freedom. It is worth noting that
first principle calculations \cite{Pecharsky2003a,Samolyuk2005}
have shown that exchange coupling is higher in the orthorhombic
phase, which stabilizes that phase at low temperatures. Also, for
Gd$_5$Ge$_4$, it was shown  \cite{Pecharsky2003b}that magnetic
degrees of freedom play a primary role in driving the
magnetostructural transition in that compound.

From the above arguments concerning crystallographic changes, the
properties of the forward martensitic transition must be compared
with those of the reverse magnetostructural transition, and
vice-versa. Two independent experimental findings are consistent
with this point of view.  The first is the fact that the amount of
AE across the forward magnetostructural transition is greater than
across the reverse transition, while, typically for martensitic
transitions, AE is greater during the reverse transition
\cite{Manosa1989b}.  The second finding is concerned with
nucleation.  According to TEM experiments \cite{Meyers2003} no
evidence of nucleation of the orthorhombic phase was observed on
cooling Gd$_5$Si$_2$Ge$_2$.  This observation is consistent with
the fact that there is no nucleation for the reverse (low-symmetry
to high symmetry) martensitic transition.

Our results show that acoustic activity evolves towards a more
reproducible pattern with thermal cycling through the
magnetostructural transition.  Actually, some dependence on thermal
cycling has been previously reported in resistance
\cite{Levin2001,Sousa2003}, thermopower \cite{Sousa2002}, and
calorimetric measurements \cite{Casanova2005APL} in
Gd$_5$(Si$_x$Ge$_{1-x}$)$_4$ alloys.  Evolution with thermal cycling
is frequently observed in first-order structural phase transitions as,
for instance, martensitic transformations \cite{PerezReche2004}. Such
evolution has been interpreted as a learning process in which the
system seeks an optimal path for the transition that tends to reduce
the dissipated energy and therefore the width of the hysteresis loop.
During this learning process the characteristics of the transition
evolve towards a reproducible pattern and the entropy change
decreases.  In Gd$_5$(Si$_x$Ge$_{1-x}$)$_4$ such evolution has been
associated with the creation of microcracks during cycling
\cite{Levin2001}.  This interpretation is, in fact, analogous to the
creation of defects (mainly dislocations) during martensitic
transition.  From these arguments it seems reasonable to associate the
evolution in Gd$_5$Si$_2$Ge$_2$ to the structural degrees of freedom
involved in the transition.

AE is a very sensitive technique to study the kinetics of a structural
transition.  Present results show that the temperature-driven
magnetostructural transition in Gd$_5$Si$_2$Ge$_2$ proceeds by
multiple steps (avalanches) joining a series of metastable states,
until the system is fully transformed.  This behaviour has been
encountered in a wide variety of driven processes in different systems
(reversal of the magnetization both in standard metamagnetic
transitions \cite{Durin2004} and magnetostructural transitions
\cite{Hardy2004PowerLaw}, martensitic transition
\cite{Vives1994a,PerezReche2004c}, emergence of vortices in
superconductors \cite{Altshuler2004}, vapor condensation in porous
media \cite{Lilly1993}...).  While it is not obvious that these
systems should show similar properties, they share a common
characteristic of being spatially extended, with a complex free energy
landscape.  Quenched-in disorder is at the origin of such a complex
landscape.  For all these systems, there is a lack of characteristic
duration and size scales of these avalanches, as reflected by
power-law distributions.  As illustrated in Fig.~\ref{Fig3}, the
amplitude of AE signals recorded here follows a power law
distribution, thus showing that the magnetostructural transition in
Gd$_5$Si$_2$Ge$_2$ belongs to such a general class of systems.

\begin{figure}
\begin{center}
  \epsfig{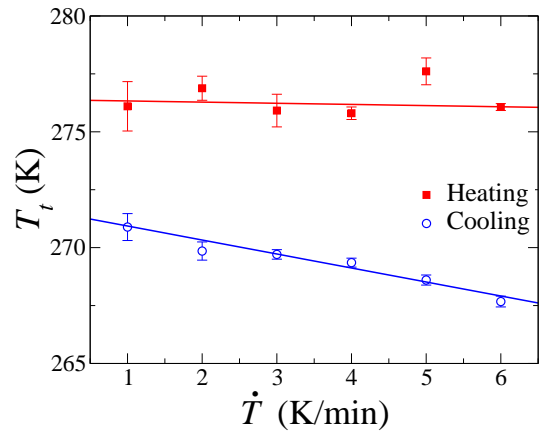}
\caption{\label{Fig4}  Average values  of the  transition temperatures
  according to Eq.~(\ref{Eq.5}) for cooling and heating runs as a
  function of $\dot{T}$.  Data corresponding to cooling and heating
  runs are indicated by different symbols according to the legend.
  Continuous lines are linear fits to the data.}
\end{center}
\end{figure}

The kinetics of these avalanche-mediated, first-order phase
transitions has been acknowledged to be athermal (i.e. thermal
fluctuations are not dominant when driving the system from one phase
to another).  The effect of thermal fluctuations (athermal degree) and
also the effect of driving rate on the power-law exponents has been
explained in terms of the different time scales involved in driving
the transition \cite{PerezReche2001,PerezReche2004c}.  Comparison of
the amplitude distributions obtained for $\dot{T}$ = 1 K/min and
$\dot{T}$ = 6 K/min shows that the power-law exponent for cooling runs
increases for increasing temperature rate. This is the behaviour
expected for those transitions for which the characteristic time of
thermal fluctuations is not very different from the characteristic
time associated with the change of the external field (temperature in
the present case).  In order to provide further support to this
statement, we measured the transition temperature as a function of the
temperature rate. For those transitions in which thermal fluctuation
cannot be neglected, it is expected that the transition temperature
decreases on increasing the temperature rate.  In contrast, no
dependence is expected for athermal transitions.  In Fig.~\ref{Fig4}
we present the average transition temperature $T_t$ for forward and
reverse magnetostructural transitions as a function of the temperature
rate. Each data point in the figure corresponds to an average over 10
cycles for each driving rate.  Due to the weakness of the detected AE,
determination of the transition temperature is not straightforward and
we have estimated it as a weighted average of $T$ with the reduced
counting rate $r$ as the weighting factor:
\begin{equation}
\label{Eq.5}
T_t=\frac{\int_{T_i}^{T_f} T r(T) dT}{\int_{T_i}^{T_f} r(T) dT},
\end{equation}
As shown in Fig.~\ref{Fig4}, a clear decrease of the transition
temperature for a monoclinic to orthorhombic transition is observed
while the orthorhombic to monoclinic transition shows no significant
dependence.  The fact that thermal fluctuations are more relevant for
the forward transition than for the reverse one is also consistent
with the results presented in Fig.~\ref{Fig2}.  Athermal transitions
are expected to be much more reproducible than those which are
affected by thermal fluctuations.

Our results are consistent with the finding of thermally activated
effects observed in the paramagnetic to ferromagnetic field-induced
transition in Gd$_5$Si$_2$Ge$_2$ \cite{Leib2004}. In this case the
application of a magnetic field induces a sudden increase in
magnetization followed by a smooth thermally activated relaxation
towards a saturation value.  Since our experiments indicate a more
athermal character on heating, we suggest that the relaxation process
observed after a demagnetizing process (ferromagnetic-monoclinic
$\rightarrow$ paramagnetic-orthorhombic) should be slower than that
observed when the magnetic field increases.

\section{CONCLUSION}

We have demonstrated the existence of acoustic emission across the
magnetostructural transition in Gd$_5$Si$_2$Ge$_2$.  The
burst-like character of the recorded AE reflects the jerky
character of the transition.  This evidences the fact that the
magnetostructural transition in giant magnetocaloric materials
proceeds by avalanches between metastable states. Moreover, there
is no characteristic scale for the size of these avalanches. The
experimental results also show that thermal fluctuations do not
play a relevant role (athermal character) for the orthorhombic
$\rightarrow$ monoclinic transition.

\section{Acknowledgements}

This work has received financial support from CICyT (Spain),
projects MAT2004-1291 and MAT2003-01124 and DURSI (Catalonia),
project 2001SGR00066.


\begin{thebibliography}{36}
\expandafter\ifx\csname
natexlab\endcsname\relax\def\natexlab#1{#1}\fi
\expandafter\ifx\csname bibnamefont\endcsname\relax
  \def\bibnamefont#1{#1}\fi
\expandafter\ifx\csname bibfnamefont\endcsname\relax
  \def\bibfnamefont#1{#1}\fi
\expandafter\ifx\csname citenamefont\endcsname\relax
  \def\citenamefont#1{#1}\fi
\expandafter\ifx\csname url\endcsname\relax
  \def\url#1{\texttt{#1}}\fi
\expandafter\ifx\csname
urlprefix\endcsname\relax\def\urlprefix{URL }\fi
\providecommand{\bibinfo}[2]{#2}
\providecommand{\eprint}[2][]{\url{#2}}

\bibitem[{\citenamefont{Planes et~al.}(2005)\citenamefont{Planes, Ma{\~n}osa,
  and Saxena}}]{Planes2005}
\bibinfo{author}{\bibfnamefont{A.}~\bibnamefont{Planes}},
  \bibinfo{author}{\bibfnamefont{L.}~\bibnamefont{Ma{\~n}osa}},
  \bibnamefont{and} \bibinfo{author}{\bibfnamefont{A.}~\bibnamefont{Saxena}},
  \emph{\bibinfo{title}{Interplay of Magnetism and Structure in Functional
  Materials}}, vol.~\bibinfo{volume}{79} (\bibinfo{publisher}{Springer-Verlag},
  \bibinfo{address}{Berlin, DE}, \bibinfo{year}{2005}).

\bibitem[{\citenamefont{K.{A.~Gschneidner,~Jr.}
  et~al.}(2005)\citenamefont{K.{A.~Gschneidner,~Jr.}, V.{K.~Pecharsky}, and
  A.{O.~Tsokol}}}]{Gschneidner2005}
\bibinfo{author}{\bibnamefont{K.{A.~Gschneidner,~Jr.}}},
  \bibinfo{author}{\bibnamefont{V.{K.~Pecharsky}}}, \bibnamefont{and}
  \bibinfo{author}{\bibnamefont{A.{O.~Tsokol}}}, \bibinfo{journal}{Rep. Prog.
  Phys.} \textbf{\bibinfo{volume}{68}}, \bibinfo{pages}{1479}
  (\bibinfo{year}{2005}).

\bibitem[{\citenamefont{Tegus et~al.}(2002)\citenamefont{Tegus, Br\"uck,
  L.Zhang, {H.J. Buschow}, and F.{R. de Boer}}}]{Tegus2002}
\bibinfo{author}{\bibfnamefont{O.}~\bibnamefont{Tegus}},
  \bibinfo{author}{\bibfnamefont{E.}~\bibnamefont{Br\"uck}},
  \bibinfo{author}{\bibnamefont{L.Zhang}},
  \bibinfo{author}{\bibfnamefont{K.}~\bibnamefont{{H.J. Buschow}}},
  \bibnamefont{and} \bibinfo{author}{\bibnamefont{F.{R. de Boer}}},
  \bibinfo{journal}{Physica B} \textbf{\bibinfo{volume}{319}},
  \bibinfo{pages}{174} (\bibinfo{year}{2002}).

\bibitem[{\citenamefont{V.{K. Pecharsky} and K.{A. Gschneider
  Jr.}}(1997)}]{Pecharsky1997}
\bibinfo{author}{\bibnamefont{V.{K. Pecharsky}}} \bibnamefont{and}
  \bibinfo{author}{\bibnamefont{K.{A. Gschneider Jr.}}},
  \bibinfo{journal}{Phys. Rev. Lett.} \textbf{\bibinfo{volume}{78}},
  \bibinfo{pages}{4494} (\bibinfo{year}{1997}).

\bibitem[{\citenamefont{Provenzano et~al.}(2004)\citenamefont{Provenzano, {J.
  Shapiro}, and {D. Schull}}}]{Provenzano2004}
\bibinfo{author}{\bibfnamefont{V.}~\bibnamefont{Provenzano}},
  \bibinfo{author}{\bibfnamefont{A.}~\bibnamefont{{J. Shapiro}}},
  \bibnamefont{and} \bibinfo{author}{\bibfnamefont{R.}~\bibnamefont{{D.
  Schull}}}, \bibinfo{journal}{Nature} \textbf{\bibinfo{volume}{429}},
  \bibinfo{pages}{853} (\bibinfo{year}{2004}).

\bibitem[{\citenamefont{Choe et~al.}(2000)\citenamefont{Choe, V.{K.~Pecharsky},
  A.{O.~Pecharsky}, K.{A.~Gschneider,~Jr.}, V.{G.~Young,~Jr.}, and
  G.{J.~Miller}}}]{Choe2000}
\bibinfo{author}{\bibfnamefont{W.}~\bibnamefont{Choe}},
  \bibinfo{author}{\bibnamefont{V.{K.~Pecharsky}}},
  \bibinfo{author}{\bibnamefont{A.{O.~Pecharsky}}},
  \bibinfo{author}{\bibnamefont{K.{A.~Gschneider,~Jr.}}},
  \bibinfo{author}{\bibnamefont{V.{G.~Young,~Jr.}}}, \bibnamefont{and}
  \bibinfo{author}{\bibnamefont{G.{J.~Miller}}}, \bibinfo{journal}{Phys. Rev.
  Lett.} \textbf{\bibinfo{volume}{84}}, \bibinfo{pages}{4617}
  (\bibinfo{year}{2000}).

\bibitem[{\citenamefont{Morellon et~al.}(1998)\citenamefont{Morellon, P.{A.
  Algarabel}, M.{R. Ibarra}, Blasco, Garc\'{\i}a-Landa, Arnold, and
  Albertini}}]{Morellon1998}
\bibinfo{author}{\bibfnamefont{L.}~\bibnamefont{Morellon}},
  \bibinfo{author}{\bibnamefont{P.{A. Algarabel}}},
  \bibinfo{author}{\bibnamefont{M.{R. Ibarra}}},
  \bibinfo{author}{\bibfnamefont{J.}~\bibnamefont{Blasco}},
  \bibinfo{author}{\bibfnamefont{B.}~\bibnamefont{Garc\'{\i}a-Landa}},
  \bibinfo{author}{\bibfnamefont{Z.}~\bibnamefont{Arnold}}, \bibnamefont{and}
  \bibinfo{author}{\bibfnamefont{F.}~\bibnamefont{Albertini}},
  \bibinfo{journal}{Phys. Rev. B} \textbf{\bibinfo{volume}{58}},
  \bibinfo{pages}{R14721} (\bibinfo{year}{1998}).

\bibitem[{\citenamefont{Mudryk et~al.}(2005)\citenamefont{Mudryk, Lee, {A.
  Gschneider, Jr.}, and {K. Pecharsky}}}]{Mudryk2005}
\bibinfo{author}{\bibfnamefont{Y.}~\bibnamefont{Mudryk}},
  \bibinfo{author}{\bibfnamefont{Y.}~\bibnamefont{Lee}},
  \bibinfo{author}{\bibfnamefont{K.}~\bibnamefont{{A. Gschneider, Jr.}}},
  \bibnamefont{and} \bibinfo{author}{\bibfnamefont{V.}~\bibnamefont{{K.
  Pecharsky}}}, \bibinfo{journal}{Phys. Rev. B} \textbf{\bibinfo{volume}{71}},
  \bibinfo{pages}{174104} (\bibinfo{year}{2005}).

\bibitem[{\citenamefont{Morellon et~al.}(2004)\citenamefont{Morellon, Arnold,
  P.{A. Algarabel}, Magen, M.{R. Ibarra}, and Skorokhod}}]{Morellon2004}
\bibinfo{author}{\bibfnamefont{L.}~\bibnamefont{Morellon}},
  \bibinfo{author}{\bibfnamefont{Z.}~\bibnamefont{Arnold}},
  \bibinfo{author}{\bibnamefont{P.{A. Algarabel}}},
  \bibinfo{author}{\bibfnamefont{C.}~\bibnamefont{Magen}},
  \bibinfo{author}{\bibnamefont{M.{R. Ibarra}}}, \bibnamefont{and}
  \bibinfo{author}{\bibfnamefont{Y.}~\bibnamefont{Skorokhod}},
  \bibinfo{journal}{J. Phys: Condens. Matter} \textbf{\bibinfo{volume}{16}},
  \bibinfo{pages}{1623} (\bibinfo{year}{2004}).

\bibitem[{\citenamefont{Casanova et~al.}(2004)\citenamefont{Casanova, Labarta,
  Batlle, Marcos, Vives, Ma{\~n}osa, Planes, and {de Brion}}}]{Casanova2004}
\bibinfo{author}{\bibfnamefont{F.}~\bibnamefont{Casanova}},
  \bibinfo{author}{\bibfnamefont{A.}~\bibnamefont{Labarta}},
  \bibinfo{author}{\bibfnamefont{X.}~\bibnamefont{Batlle}},
  \bibinfo{author}{\bibfnamefont{J.}~\bibnamefont{Marcos}},
  \bibinfo{author}{\bibfnamefont{E.}~\bibnamefont{Vives}},
  \bibinfo{author}{\bibfnamefont{L.}~\bibnamefont{Ma{\~n}osa}},
  \bibinfo{author}{\bibfnamefont{A.}~\bibnamefont{Planes}}, \bibnamefont{and}
  \bibinfo{author}{\bibfnamefont{S.}~\bibnamefont{{de Brion}}},
  \bibinfo{journal}{Phys. Rev. B} \textbf{\bibinfo{volume}{69}},
  \bibinfo{pages}{104416} (\bibinfo{year}{2004}).

\bibitem[{\citenamefont{Morellon et~al.}(2000)\citenamefont{Morellon, Blasco,
  P.{A. Algarabel}, and M.{R. Ibarra}}}]{Morellon2000}
\bibinfo{author}{\bibfnamefont{L.}~\bibnamefont{Morellon}},
  \bibinfo{author}{\bibfnamefont{J.}~\bibnamefont{Blasco}},
  \bibinfo{author}{\bibnamefont{P.{A. Algarabel}}}, \bibnamefont{and}
  \bibinfo{author}{\bibnamefont{M.{R. Ibarra}}}, \bibinfo{journal}{Phys. Rev.
  B} \textbf{\bibinfo{volume}{62}}, \bibinfo{pages}{1022}
  (\bibinfo{year}{2000}).

\bibitem[{\citenamefont{Ma{\~n}osa
  et~al.}(1989{\natexlab{a}})\citenamefont{Ma{\~n}osa, Planes, Rouby, Morin,
  Fleischman, and J.{L.~Macqueron}}}]{Manosa1989}
\bibinfo{author}{\bibfnamefont{L.}~\bibnamefont{Ma{\~n}osa}},
  \bibinfo{author}{\bibfnamefont{A.}~\bibnamefont{Planes}},
  \bibinfo{author}{\bibfnamefont{D.}~\bibnamefont{Rouby}},
  \bibinfo{author}{\bibfnamefont{M.}~\bibnamefont{Morin}},
  \bibinfo{author}{\bibfnamefont{P.}~\bibnamefont{Fleischman}},
  \bibnamefont{and} \bibinfo{author}{\bibnamefont{J.{L.~Macqueron}}},
  \bibinfo{journal}{Appl. Phys. Lett.} \textbf{\bibinfo{volume}{54}},
  \bibinfo{pages}{2574} (\bibinfo{year}{1989}{\natexlab{a}}).

\bibitem[{\citenamefont{Yu and P.{C.~Clapp}}(1987)}]{Yu1987}
\bibinfo{author}{\bibfnamefont{Z.}~\bibnamefont{Yu}} \bibnamefont{and}
  \bibinfo{author}{\bibnamefont{P.{C.~Clapp}}}, \bibinfo{journal}{J. Appl.
  Phys.} \textbf{\bibinfo{volume}{62}}, \bibinfo{pages}{2212 }
  (\bibinfo{year}{1987}).

\bibitem[{\citenamefont{Vives et~al.}(1994)\citenamefont{Vives, Ort\'{\i}n,
  Ma{\~n}osa, R\`{a}fols, P\'{e}rez-Magran\'{e}, and Planes}}]{Vives1994a}
\bibinfo{author}{\bibfnamefont{E.}~\bibnamefont{Vives}},
  \bibinfo{author}{\bibfnamefont{J.}~\bibnamefont{Ort\'{\i}n}},
  \bibinfo{author}{\bibfnamefont{L.}~\bibnamefont{Ma{\~n}osa}},
  \bibinfo{author}{\bibfnamefont{I.}~\bibnamefont{R\`{a}fols}},
  \bibinfo{author}{\bibfnamefont{R.}~\bibnamefont{P\'{e}rez-Magran\'{e}}},
  \bibnamefont{and} \bibinfo{author}{\bibfnamefont{A.}~\bibnamefont{Planes}},
  \bibinfo{journal}{Phys.\ Rev.\ Lett.} \textbf{\bibinfo{volume}{72}},
  \bibinfo{pages}{1694} (\bibinfo{year}{1994}).

\bibitem[{\citenamefont{F.{J.~P\'erez-Reche}
  et~al.}(2001)\citenamefont{F.{J.~P\'erez-Reche}, Vives, Ma{\~n}osa, and
  Planes}}]{PerezReche2001}
\bibinfo{author}{\bibnamefont{F.{J.~P\'erez-Reche}}},
  \bibinfo{author}{\bibfnamefont{E.}~\bibnamefont{Vives}},
  \bibinfo{author}{\bibfnamefont{L.}~\bibnamefont{Ma{\~n}osa}},
  \bibnamefont{and} \bibinfo{author}{\bibfnamefont{A.}~\bibnamefont{Planes}},
  \bibinfo{journal}{Phys. Rev. Lett.} \textbf{\bibinfo{volume}{87}},
  \bibinfo{pages}{195701} (\bibinfo{year}{2001}).

\bibitem[{\citenamefont{Ma{\~n}osa et~al.}(1990)\citenamefont{Ma{\~n}osa,
  Planes, Rouby, and J.{L.~Macqueron}}}]{Manosa1990}
\bibinfo{author}{\bibfnamefont{L.}~\bibnamefont{Ma{\~n}osa}},
  \bibinfo{author}{\bibfnamefont{A.}~\bibnamefont{Planes}},
  \bibinfo{author}{\bibfnamefont{D.}~\bibnamefont{Rouby}}, \bibnamefont{and}
  \bibinfo{author}{\bibnamefont{J.{L.~Macqueron}}}, \bibinfo{journal}{Acta
  Metall. Mater.} \textbf{\bibinfo{volume}{38}}, \bibinfo{pages}{1635}
  (\bibinfo{year}{1990}).

\bibitem[{\citenamefont{J.{S.~Meyers} et~al.}(2003)\citenamefont{J.{S.~Meyers},
  Chumbley, Laabs, and A.{O.~Pecharsky}}}]{Meyers2003}
\bibinfo{author}{\bibnamefont{J.{S.~Meyers}}},
  \bibinfo{author}{\bibfnamefont{S.}~\bibnamefont{Chumbley}},
  \bibinfo{author}{\bibfnamefont{F.}~\bibnamefont{Laabs}}, \bibnamefont{and}
  \bibinfo{author}{\bibnamefont{A.{O.~Pecharsky}}}, \bibinfo{journal}{Acta
  Mater.} \textbf{\bibinfo{volume}{51}}, \bibinfo{pages}{61}
  (\bibinfo{year}{2003}).

\bibitem[{\citenamefont{Pecharsky and Jr}(1997)}]{Pecharsky1997b}
\bibinfo{author}{\bibfnamefont{V.}~\bibnamefont{Pecharsky}} \bibnamefont{and}
  \bibinfo{author}{\bibfnamefont{K.~G.} \bibnamefont{Jr}}, \bibinfo{journal}{J.
  Alloys Compd.} \textbf{\bibinfo{volume}{260}}, \bibinfo{pages}{98}
  (\bibinfo{year}{1997}).

\bibitem[{\citenamefont{E.{M.~Levin}
  et~al.}(2001{\natexlab{a}})\citenamefont{E.{M.~Levin}, V.{K.~Pecharsky}, and
  K.{A.~Gschneidner,~Jr.}}}]{Levin2001b}
\bibinfo{author}{\bibnamefont{E.{M.~Levin}}},
  \bibinfo{author}{\bibnamefont{V.{K.~Pecharsky}}}, \bibnamefont{and}
  \bibinfo{author}{\bibnamefont{K.{A.~Gschneidner,~Jr.}}},
  \bibinfo{journal}{Phys. Rev. B} \textbf{\bibinfo{volume}{63}},
  \bibinfo{pages}{174110} (\bibinfo{year}{2001}{\natexlab{a}}).

\bibitem[{\citenamefont{Ma{\~n}osa
  et~al.}(1989{\natexlab{b}})\citenamefont{Ma{\~n}osa, Planes, and
  Cesari}}]{Manosa1989b}
\bibinfo{author}{\bibfnamefont{L.}~\bibnamefont{Ma{\~n}osa}},
  \bibinfo{author}{\bibfnamefont{A.}~\bibnamefont{Planes}}, \bibnamefont{and}
  \bibinfo{author}{\bibfnamefont{E.}~\bibnamefont{Cesari}},
  \bibinfo{journal}{J. Phys. D:Appl. Phys.} \textbf{\bibinfo{volume}{22}},
  \bibinfo{pages}{977} (\bibinfo{year}{1989}{\natexlab{b}}).

\bibitem[{\citenamefont{F.{J.~P\'erez-Reche}
  et~al.}(2004{\natexlab{a}})\citenamefont{F.{J.~P\'erez-Reche}, Stipcich,
  Vives, Ma{\~n}osa, Planes, and Morin}}]{PerezReche2004}
\bibinfo{author}{\bibnamefont{F.{J.~P\'erez-Reche}}},
  \bibinfo{author}{\bibfnamefont{M.}~\bibnamefont{Stipcich}},
  \bibinfo{author}{\bibfnamefont{E.}~\bibnamefont{Vives}},
  \bibinfo{author}{\bibfnamefont{L.}~\bibnamefont{Ma{\~n}osa}},
  \bibinfo{author}{\bibfnamefont{A.}~\bibnamefont{Planes}}, \bibnamefont{and}
  \bibinfo{author}{\bibfnamefont{M.}~\bibnamefont{Morin}},
  \bibinfo{journal}{Phys. Rev. B} \textbf{\bibinfo{volume}{69}},
  \bibinfo{pages}{064101} (\bibinfo{year}{2004}{\natexlab{a}}).

\bibitem[{\citenamefont{M.{L.~Goldstein}
  et~al.}(2004)\citenamefont{M.{L.~Goldstein}, S.{A.~Morris}, and
  G.{G.~Yen}}}]{Goldstein2004}
\bibinfo{author}{\bibnamefont{M.{L.~Goldstein}}},
  \bibinfo{author}{\bibnamefont{S.{A.~Morris}}}, \bibnamefont{and}
  \bibinfo{author}{\bibnamefont{G.{G.~Yen}}}, \bibinfo{journal}{Eur. Phys. J.
  B} \textbf{\bibinfo{volume}{41}}, \bibinfo{pages}{255 }
  (\bibinfo{year}{2004}).

\bibitem[{\citenamefont{Gu{\'e}nin and P.{F.~Gobin}}(1982)}]{Guenin1982}
\bibinfo{author}{\bibfnamefont{G.}~\bibnamefont{Gu{\'e}nin}} \bibnamefont{and}
  \bibinfo{author}{\bibnamefont{P.{F.~Gobin}}}, \bibinfo{journal}{J. Phys
  (Paris)} \textbf{\bibinfo{volume}{43}}, \bibinfo{pages}{C4}
  (\bibinfo{year}{1982}).

\bibitem[{\citenamefont{Samolyuk and Antropov}(2005)}]{Samolyuk2005}
\bibinfo{author}{\bibfnamefont{G.}~\bibnamefont{Samolyuk}} \bibnamefont{and}
  \bibinfo{author}{\bibfnamefont{V.}~\bibnamefont{Antropov}},
  \bibinfo{journal}{J. Appl. Phys.} \textbf{\bibinfo{volume}{97}},
  \bibinfo{pages}{10A310} (\bibinfo{year}{2005}).

\bibitem[{\citenamefont{V.{K.~Pecharsky}
  et~al.}(2003{\natexlab{a}})\citenamefont{V.{K.~Pecharsky},
  K.{A.~Gschneidner,~Jr.}, G.{D. Samolyuk}, V.{P. Antropov}, A.{O. Pecharsky},
  and K.{A.~Gschneidner,~Jr.}}}]{Pecharsky2003a}
\bibinfo{author}{\bibnamefont{V.{K.~Pecharsky}}},
  \bibinfo{author}{\bibnamefont{K.{A.~Gschneidner,~Jr.}}},
  \bibinfo{author}{\bibnamefont{G.{D. Samolyuk}}},
  \bibinfo{author}{\bibnamefont{V.{P. Antropov}}},
  \bibinfo{author}{\bibnamefont{A.{O. Pecharsky}}}, \bibnamefont{and}
  \bibinfo{author}{\bibnamefont{K.{A.~Gschneidner,~Jr.}}}, \bibinfo{journal}{J.
  Solid State Chem.} \textbf{\bibinfo{volume}{171}}, \bibinfo{pages}{57}
  (\bibinfo{year}{2003}{\natexlab{a}}).

\bibitem[{\citenamefont{V.{K.~Pecharsky}
  et~al.}(2003{\natexlab{b}})\citenamefont{V.{K.~Pecharsky}, A.{P.Holm},
  K.{A.~Gschneidner,~Jr.}, and Rink}}]{Pecharsky2003b}
\bibinfo{author}{\bibnamefont{V.{K.~Pecharsky}}},
  \bibinfo{author}{\bibnamefont{A.{P.Holm}}},
  \bibinfo{author}{\bibnamefont{K.{A.~Gschneidner,~Jr.}}}, \bibnamefont{and}
  \bibinfo{author}{\bibfnamefont{R.}~\bibnamefont{Rink}},
  \bibinfo{journal}{Phys. Rev. Lett.} \textbf{\bibinfo{volume}{91}},
  \bibinfo{pages}{197204} (\bibinfo{year}{2003}{\natexlab{b}}).

\bibitem[{\citenamefont{E.{M.~Levin}
  et~al.}(2001{\natexlab{b}})\citenamefont{E.{M.~Levin}, A.{O.~Pecharsky},
  V.{K.~Pecharsky}, and K.{A.~Gschneidner,~Jr.}}}]{Levin2001}
\bibinfo{author}{\bibnamefont{E.{M.~Levin}}},
  \bibinfo{author}{\bibnamefont{A.{O.~Pecharsky}}},
  \bibinfo{author}{\bibnamefont{V.{K.~Pecharsky}}}, \bibnamefont{and}
  \bibinfo{author}{\bibnamefont{K.{A.~Gschneidner,~Jr.}}},
  \bibinfo{journal}{Phys. Rev. B} \textbf{\bibinfo{volume}{63}},
  \bibinfo{pages}{064426} (\bibinfo{year}{2001}{\natexlab{b}}).

\bibitem[{\citenamefont{J.{B.~Sousa} et~al.}(2003)\citenamefont{J.{B.~Sousa},
  M.{E.~Braga}, F.{C.~Correia}, Carpinteiro, Morellon, P.{A.~Algarabel}, and
  M.{R.~Ibarra}}}]{Sousa2003}
\bibinfo{author}{\bibnamefont{J.{B.~Sousa}}},
  \bibinfo{author}{\bibnamefont{M.{E.~Braga}}},
  \bibinfo{author}{\bibnamefont{F.{C.~Correia}}},
  \bibinfo{author}{\bibfnamefont{F.}~\bibnamefont{Carpinteiro}},
  \bibinfo{author}{\bibfnamefont{L.}~\bibnamefont{Morellon}},
  \bibinfo{author}{\bibnamefont{P.{A.~Algarabel}}}, \bibnamefont{and}
  \bibinfo{author}{\bibnamefont{M.{R.~Ibarra}}}, \bibinfo{journal}{Phys. Rev.
  B} \textbf{\bibinfo{volume}{67}}, \bibinfo{pages}{134416}
  (\bibinfo{year}{2003}).

\bibitem[{\citenamefont{J.{B.~Sousa} et~al.}(2002)\citenamefont{J.{B.~Sousa},
  M.{E.~Braga}, F.{C.~Correira}, Carpinteiro, Morellon, P.{A.~Algarabel}, and
  M.{R.~Ibarra}}}]{Sousa2002}
\bibinfo{author}{\bibnamefont{J.{B.~Sousa}}},
  \bibinfo{author}{\bibnamefont{M.{E.~Braga}}},
  \bibinfo{author}{\bibnamefont{F.{C.~Correira}}},
  \bibinfo{author}{\bibfnamefont{F.}~\bibnamefont{Carpinteiro}},
  \bibinfo{author}{\bibfnamefont{L.}~\bibnamefont{Morellon}},
  \bibinfo{author}{\bibnamefont{P.{A.~Algarabel}}}, \bibnamefont{and}
  \bibinfo{author}{\bibnamefont{M.{R.~Ibarra}}}, \bibinfo{journal}{J. Appl.
  Phys.} \textbf{\bibinfo{volume}{91}}, \bibinfo{pages}{4457}
  (\bibinfo{year}{2002}).

\bibitem[{\citenamefont{Casanova et~al.}(2005)\citenamefont{Casanova, Labarta,
  Batlle, F.{J.~P\'erez-Reche}, Vives, Ma{\~n}osa, and
  Planes}}]{Casanova2005APL}
\bibinfo{author}{\bibfnamefont{F.}~\bibnamefont{Casanova}},
  \bibinfo{author}{\bibfnamefont{A.}~\bibnamefont{Labarta}},
  \bibinfo{author}{\bibfnamefont{X.}~\bibnamefont{Batlle}},
  \bibinfo{author}{\bibnamefont{F.{J.~P\'erez-Reche}}},
  \bibinfo{author}{\bibfnamefont{E.}~\bibnamefont{Vives}},
  \bibinfo{author}{\bibfnamefont{L.}~\bibnamefont{Ma{\~n}osa}},
  \bibnamefont{and} \bibinfo{author}{\bibfnamefont{A.}~\bibnamefont{Planes}},
  \bibinfo{journal}{Appl. Phys. Lett.} \textbf{\bibinfo{volume}{86}},
  \bibinfo{pages}{262504} (\bibinfo{year}{2005}).

\bibitem[{\citenamefont{Durin and Zapperi}(2004)}]{Durin2004}
\bibinfo{author}{\bibfnamefont{G.}~\bibnamefont{Durin}} \bibnamefont{and}
  \bibinfo{author}{\bibfnamefont{S.}~\bibnamefont{Zapperi}},
  \bibinfo{journal}{cond-mat/0404512}  (\bibinfo{year}{2004}).

\bibitem[{\citenamefont{Hardy et~al.}(2004)\citenamefont{Hardy, Majumdar,
  M.{R.~Lees}, D.{~McK.~Paul}, Yaicle, and Hervieu}}]{Hardy2004PowerLaw}
\bibinfo{author}{\bibfnamefont{V.}~\bibnamefont{Hardy}},
  \bibinfo{author}{\bibfnamefont{S.}~\bibnamefont{Majumdar}},
  \bibinfo{author}{\bibnamefont{M.{R.~Lees}}},
  \bibinfo{author}{\bibnamefont{D.{~McK.~Paul}}},
  \bibinfo{author}{\bibfnamefont{C.}~\bibnamefont{Yaicle}}, \bibnamefont{and}
  \bibinfo{author}{\bibfnamefont{M.}~\bibnamefont{Hervieu}},
  \bibinfo{journal}{Phys. Rev. B} \textbf{\bibinfo{volume}{70}},
  \bibinfo{pages}{104423} (\bibinfo{year}{2004}).

\bibitem[{\citenamefont{F.{J.~P\'erez-Reche}
  et~al.}(2004{\natexlab{b}})\citenamefont{F.{J.~P\'erez-Reche}, Tadi{\'c},
  Ma{\~n}osa, Planes, and Vives}}]{PerezReche2004c}
\bibinfo{author}{\bibnamefont{F.{J.~P\'erez-Reche}}},
  \bibinfo{author}{\bibfnamefont{B.}~\bibnamefont{Tadi{\'c}}},
  \bibinfo{author}{\bibfnamefont{L.}~\bibnamefont{Ma{\~n}osa}},
  \bibinfo{author}{\bibfnamefont{A.}~\bibnamefont{Planes}}, \bibnamefont{and}
  \bibinfo{author}{\bibfnamefont{E.}~\bibnamefont{Vives}},
  \bibinfo{journal}{Phys. Rev. Lett.} \textbf{\bibinfo{volume}{93}},
  \bibinfo{pages}{195701} (\bibinfo{year}{2004}{\natexlab{b}}).

\bibitem[{\citenamefont{Altshuler and T.{H.~Johansen}}(2004)}]{Altshuler2004}
\bibinfo{author}{\bibfnamefont{E.}~\bibnamefont{Altshuler}} \bibnamefont{and}
  \bibinfo{author}{\bibnamefont{T.{H.~Johansen}}}, \bibinfo{journal}{Rev. Mod.
  Phys.} \textbf{\bibinfo{volume}{76}}, \bibinfo{pages}{471}
  (\bibinfo{year}{2004}), \bibinfo{note}{and refs. therein.}

\bibitem[{\citenamefont{M.{P.~Lilly} et~al.}(1993)\citenamefont{M.{P.~Lilly},
  P.{T.~Finley}, and R.{B.~Hallock}}}]{Lilly1993}
\bibinfo{author}{\bibnamefont{M.{P.~Lilly}}},
  \bibinfo{author}{\bibnamefont{P.{T.~Finley}}}, \bibnamefont{and}
  \bibinfo{author}{\bibnamefont{R.{B.~Hallock}}}, \bibinfo{journal}{Phys.\
  Rev.\ Lett.} \textbf{\bibinfo{volume}{71}}, \bibinfo{pages}{4186}
  (\bibinfo{year}{1993}).

\bibitem[{\citenamefont{Leib et~al.}(2004)\citenamefont{Leib, J.{E.~Snyder},
  T.{A.~Lograsso}, Schlagel, and D.{C.~Jiles}}}]{Leib2004}
\bibinfo{author}{\bibfnamefont{J.}~\bibnamefont{Leib}},
  \bibinfo{author}{\bibnamefont{J.{E.~Snyder}}},
  \bibinfo{author}{\bibnamefont{T.{A.~Lograsso}}},
  \bibinfo{author}{\bibfnamefont{D.}~\bibnamefont{Schlagel}}, \bibnamefont{and}
  \bibinfo{author}{\bibnamefont{D.{C.~Jiles}}}, \bibinfo{journal}{J. Appl.
  Phys.} \textbf{\bibinfo{volume}{95}}, \bibinfo{pages}{6915}
  (\bibinfo{year}{2004}).

\end{thebibliography}

\end{document}